\documentclass[preprint2]{aastex}
\usepackage{graphicx}
\usepackage{natbib}

\def\msun{M$_{\odot}$}

\def\mdot{$\dot M$}

\def\it{\sl}
\def\degs{\ifmmode ^{\circ}\else$^{\circ}$\fi}
\def\amin{\ifmmode ^{\prime}\else$^{\prime}$\fi}
\def\asec{\ifmmode ^{\prime\prime}\else$^{\prime\prime}$\fi}
\def\fd{\hbox{$.\!\!^{\rm d}$}}            
\def\farcs{\hbox{$.\!\!^{\prime\prime}$}}  

\def\degs{\ifmmode ^{\circ}\else$^{\circ}$\fi}
\def\amin{\ifmmode ^{\prime}\else$^{\prime}$\fi}

\def\eqalign#1{\null\,\vcenter{\openup1\jot \m@th
   \ialign{\strut\hfil$\displaystyle{##}$&$\displaystyle{{}##}$\hfil
   \crcr#1\crcr}}\,}
\usepackage{color}

\newcommand{\sdss}{SDSS\,0756+0858}

\def\apgt{\ {\raise-.5ex\hbox{$\buildrel>\over\sim$}}\ }
\def\aplt{\ {\raise-.5ex\hbox{$\buildrel<\over\sim$}}\ }

\shorttitle{On the SW Sex-Type  CV SDSS0756+0858}
\shortauthors{Tovmassian et al.}

\begin{document}

\title{On the SW Sex-Type Eclipsing Cataclysmic Variable SDSS0756+0858}

\author{Gagik~Tovmassian,
  Mercedes~Stephania~Hernandez,
  Diego~Gonz\'alez-Buitrago,
  Sergey~Zharikov
  and
  Maria~Teresa~Garc\'{i}a-D\'{i}az }

\affil{Instituto de Astronom\'{\i}a, Universidad Nacional Autonoma
de M\'exico, Apdo. Postal 877, Ensenada, Baja California, 22800 M\'exico}
\email{gag@astro.unam.mx}
\begin{abstract}
We conducted a spectroscopic and photometric study of SDSS J075653.11+085831.  X-ray observations were also attempted.  We determined the orbital period of this binary system to be  3.29 hr. It is a deep eclipsing system, whose spectra shows mostly single-peaked Balmer emission lines and a rather intense He\,{\sc ii} line. There is also the presence of faint (often double-peaked) He{\sc I} emission lines 
as well as several absorption lines, Mg\,{\sc i} being the most prominent.   All of these features  point towards the affiliation of this object with the growing number of SW~Sex-type objects. We developed a phenomenological model of an SW~Sex system to reproduce the observed photometric and spectral features.
\end{abstract}


\maketitle

\section{Introduction}
\label{sec:intro}

Cataclysmic variables (CVs) are close, interactive  binaries comprised of a white dwarf and a late-type,
main sequence star. The latter fills its corresponding Roche lobe and loses matter via the Lagrangian L$_1$ point to the more massive, compact  component. In most  cases, the transferred matter forms an accretion disk around the white dwarf if the former has no strong magnetic field. 
The observational characteristics of  CVs, upon which they are normally classified,  depend greatly on the orbital period of the system, the mass accretion rate,  and the strength of the above-mentioned  magnetic field.  The orbital period reflects not only  the separation of the binary components, but also  the size of  the Roche lobe-filling secondary and therefore  its evolutionary status.  In particular, CVs with periods over $\sim3$\,hours still contain  secondaries  with  radiative cores,  and thus  rely on magnetic breaking as a driving force to remove the angular momentum  \citep{1983ApJ...275..713R}. Systems with  periods below this 3 hr threshold  contain red dwarfs, which are fully convective.  
The magnetic breaking then becomes negligible, halting the mass transfer.  Before this happens,  a significant number of CVs  known  as  SW Sex-type are observed  within the $ 3 \sim 4$ hr period range \citep{2012arXiv1211.2171S}. 
The SW Sex stars were first identified by \citet{1991AJ....102..272T} as  eclipsing systems, which show characteristics
very similar to each other and quite different from other CVs. Apart from wide, non-symmetric eclipses, SW\,Sex stars
exhibit  high excitation lines as well as absorption features,  shifts between eclipse and emission lines phasing, and single-peaked line profiles (unusual for  eclipsing binaries)  containing accretion disks  \citep{1991MNRAS.252..342D, 2000NewAR..44..131H, 2003AJ....126.2473H, 2007MNRAS.377.1747R}. 
They are also defined as nova-likes, because the high mass transfer rate helps to maintain a hot and stable accretion disk 
devoid of dwarf-nova-type  outbursts. A detailed  description and discussion of properties of SW Sex stars can be found in a number of papers from the discovery of the phenomenon \citep{1991AJ....102..272T}  to the latest reviews \citep{2007MNRAS.377.1747R,2013MNRAS.428.3559D}.

SDSS J075653.11+085831.8 (hereinafter SDSS0756+0858) was identified as a CV by \citet{2011AJ....142..181S} from the SDSS DR7 lists. From  a very limited time coverage ($\sim1.5$\,hr)  they suggested that the object may have a period around two hours and thought it could be an intermediate polar. Since they covered slightly less than the actual orbital period of the system, they probably missed the eclipse, without which a correct identification becomes very difficult. We observed the object during two seasons in 2012 and 2013 by means of spectroscopy and multicolour photometry. We find that SDSS0756+0858 is a bona-fide member of the SW Sex stars. Here, we present the results of our observations, discuss the deduced parameters of the system and  its geometry. The observations and  data reduction are described  in Section~\ref{sec:obs}, while the  results are presented  and discussed in Sections~\ref{sec:porb},  \ref{sec:param}, and \ref{sec:conclud}.  In Section~\ref{sec:model} we present our model of SW\,Sex stars and draw brief conclusions in Section~\ref{sec:conclud2}.

\section{Observations and Data Reduction}
\label{sec:obs}

\begin{figure}[t]
\setlength{\unitlength}{1mm}
\resizebox{11cm}{!}{
\begin{picture}(100,70)(0,0)
\put (0,0)  { \includegraphics[width=7cm,  clip]{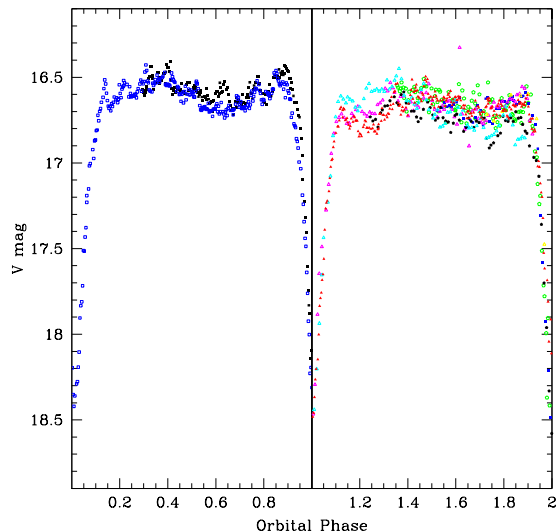}}
\end{picture}}
  \caption{$V$ band light curve of \sdss\  folded with the orbital period (see Section\,\ref{sec:porb}).   Left panel:  a few orbits of V band light curves  with distinct hump before the eclipse, and a pronounced dip around phase 0.6--0.7 are presented.  Right panel: the rest of the  light curves without  the pre-eclipse hump  and with  a depression  after the eclipse are shown.  }
  \label{fig:lc}
\end{figure}

\subsection{Photometry}
\label{sec:phot}

\begin{figure*}[t]
\setlength{\unitlength}{1mm}
\resizebox{11cm}{!}{
\begin{picture}(100,120)(0,0)
\put (0,0)  {
\includegraphics[width=15cm, bb=20 150 580 720, clip]{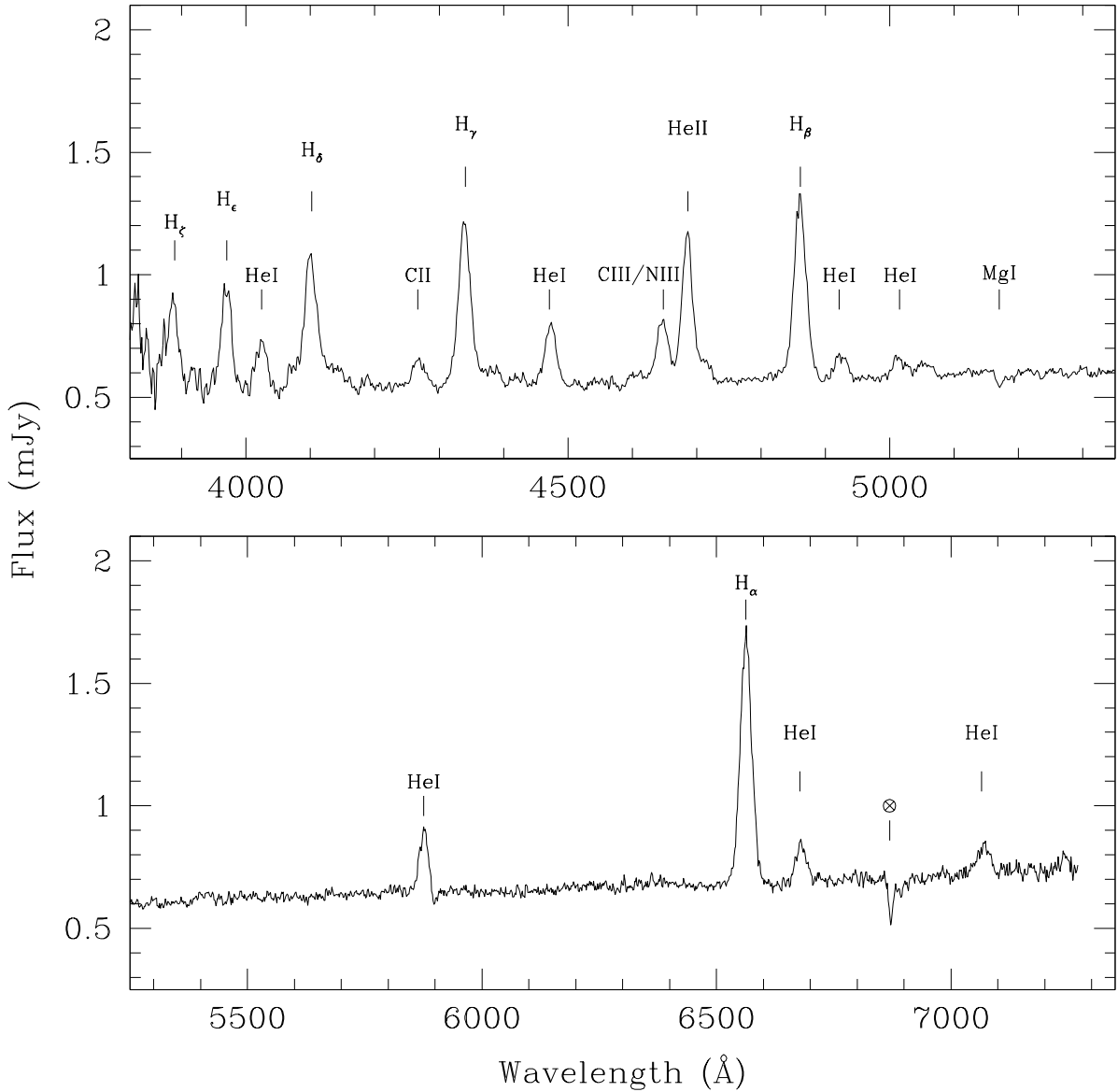}
   }
\end{picture}}
\caption{Average spectrum of \sdss.  The major lines are marked.}
  \label{fig:spec}
\end{figure*}

Time-resolved photometry  of \sdss\ was obtained using the direct  CCD imaging mode
of the 0.84\,m telescope of  the Observatorio Astron\'omico Nacional at San Pedro M\'artir
(OAN SPM)\footnote{http://www.astrossp.unam.mx} in Mexico. 
We obtained a long series of photometry in the $V$\,band and one or two orbital periods with each $UBRI$ Johnson-Cousins filter.
The log of photometric observations is given in Table~\ref{tab:log}. 
The photometric data were calibrated using Landolt standard stars. 
The errors, ranging from 0.01 to 0.05 mag, were estimated from the magnitude dispersion of  comparison stars with similar brightness. 
The  light curve shows deep ($\sim2$ mag) and wide ($\sim0.2$\,P$_{\rm {orb}}$)  eclipses. 
 Outside the eclipse, the object brightness   varies smoothly with superposed short-time flickering. The ingress to the eclipse is steeper than the egress. Such an eclipse shape is very common for SW\,Sex stars \citep{1991AJ....102..272T} and was  a defining feature of the class until it was  later revealed that there are also non-eclipsing SW\,Sex stars \citep{2007MNRAS.374.1359R}. 

In Figure\,\ref{fig:lc},  the $V$-band light curve, comprised of nine orbits  and folded with the orbital period,  is presented.  The period determination is described in the next section. Two cycles are plotted to separate two distinct light curve morphologies.
The left panel  shows two orbits  with a distinct hump before an eclipse, and a pronounced dip around phase of $\phi\sim0.6-0.7$. In  the right panel, the remaining  light curves are combined, where  the pre-eclipse hump is not so prominent and there is a distinct depression of brightness 
right after the  eclipse.  The dip around phase $\phi\sim0.6$ is also not as defined as it is in the left panel. In both cases, the object reaches maximum brightness at  phase $\phi\sim0.4$.

\subsection{Spectroscopy}
\label{sec:spec}

\begin{table}[t]
 \centering
    \caption{Log of observations} 
    \vspace{0.2cm}
\begin{tabular}{ccccc} \hline\hline
Spectrosopy$^\dagger$&   Coverage  & Resolution &  Exposure       \\    
Date & hours &  FWHM \AA & sec \\ \hline
01/02/2012   & 7.3   & 2.1 &  600-1200       \\
04/03/2012  &  5.8   & 4.4 &  600-1200     \\ 
05/12/2012  &  1.2   & 2.1 &  900               \\ 
06/12/2012  &  3.4   & 2.1 &  900                \\ 
07/12/2012  &  1.5   & 2.1 &  1200                 \\ 
16/01/2013  &  2.6   & 2.1 &  1200              \\ 
17/01/2013  &  5.4   & 2.1 &  1200             \\ 
06/02/2013  &  3.1   & 6.5 &  600                \\
07/02/2013  &  4.9   & 6.5 &  600                 \\
\hline
 \vspace{0.2cm}
Photometry   $^\dagger$$^\dagger$&   & Filter &          \\     \hline

01/02/2012  &  9    &    V            & 40            \\
03/02/2012  &  2.4 &  B/V/R/I    & 60-120    \\
04/03/2012  &  6.2 &     V           &  60            \\
05/03/2012  &  6    &     V           & 40              \\
17/01/2013  &  4.0    &     B           & 90             \\
19/01/2013  & 9.8    &     V/R           & 120        \\
20/01/2013  &  7.6    &     U/I           & 90-120    \\
06/02/2013  &  4.4    &     V           &120            \\
07/02/2013  &  5.4    &     V           &120             \\
 \hline
 \hline

\end{tabular}
\begin{tabular}{l}
$^\dagger$ {2.12\,m telescope;} 
$^\dagger$$^\dagger$ {0.84\,m telescope OAN SPM} 
\end{tabular}
\label{tab:log}
\end{table}

\begin{figure}[t]
\setlength{\unitlength}{1mm}
\resizebox{11cm}{!}{
\begin{picture}(100,70)(0,0)
\put (0,0)  {
\includegraphics[width=7cm, bb=20 145 590 715, clip]{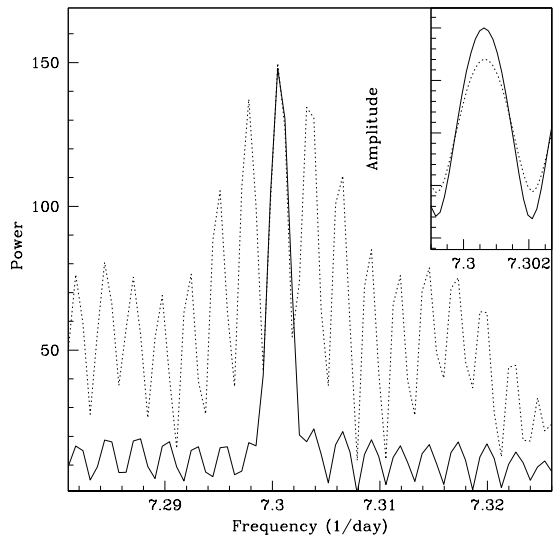}
   }
\end{picture}}
  \caption{The power spectra of \sdss, obtained from DFT analysis. The dotted line is the spectrum obtained for  H$_\beta$ radial velocities. The solid line is the same after alias frequency cleaning.
  In the inset of the figure only the  highest peak is  plotted (and over-plotted is the amplitude of the power spectrum obtained from the photometry). There is excellent coincidence of frequencies, but since the amplitudes are very different the curves were scaled to fit in the same frame.  }
  \label{fig:power}
\end{figure}

We obtained long-slit spectroscopy of \sdss\ during several observing runs using  the 2.12\,m  telescope  and the Boller~\&~Chivens spectrograph at OAN SPM.  Two different gratings were used. The 1200 {\it l/mm} grating was used  to obtain  the emission line profiles with  the highest possible spectral resolution (up to 2.1\,\AA\ FWHM  in the covered range $\sim4000 - 5200$\,\AA)  in this faint $V\sim16.25$ object. 
In addition, 6.5\,\AA\ resolution spectra with a higher S/N ratio were obtained with a 400 {\it l/mm} grating, to study the spectral energy distribution of the system, with  a broader  wavelength coverage ($\sim 4000-7200$\AA).  Observations were made with a 1\farcs5 slit, oriented in the Eeast-West direction. CuNeArg-arcs  were taken every 30-40 min for wavelength calibration. The spectrograph flexions have also been checked  with the night-sky lines.  Spectrophotometric standards from \cite{1990AJ.....99.1621O} were observed each night for flux calibration. Flux losses due to the slit width and slit orientation have not been taken into account, but in some cases we have obtained simultaneous photometry and can flux-calibrate them, if necessary.  The log of spectroscopic observations and the details of the instrumental setting are provided in Table\,\ref{tab:log}.  

The median of all 28 spectra of \sdss\ taken on 2013 Feb 7,   is presented in Figure\,\ref{fig:spec}. 
The object shows single-peaked emission lines of hydrogen (Balmer series), He\,{\sc i} lines, He\,{\sc ii} $\lambda\,4686$\,\AA, Bowen fluorescence  C\,{\sc iii}/N\,{\sc iii} $\lambda\lambda\,4640-4650$\,\AA\ and C\,{\sc ii} $\lambda\lambda4267$, 7231, 7236\,\AA\ lines. 
Only immediately before the eclipse they show a split, which is better pronounced in the H$_\alpha$ line.  He\,{\sc i} lines show transient double peaks more often. Absorption features are also transient. The  most prominent among them is the Mg line blend around $\lambda5175$\,\AA. 
 The spectra taken in different epochs did not vary much in  shape, therefore the spectrum in Figure\,\ref{fig:spec} is representative of the object. 

The parameters of spectral lines (radial velocities, intensities and FWHM) were measured by fitting  a single Lorentzian to the emission lines, which describes observed profiles of lines much better (complete) than a Gaussian profile. The reason is obvious:  the lines are formed in an extended area with different velocities and are wider and distorted from an instrumental profile. Often, broad emission lines of CVs are measured by  so-called double Gaussian method. This procedure is very useful  when dealing with the lines formed in an accretion disk and it allows one to measure velocity amplitude close to the central star. However, in this case, we do not think the lines are formed in the entire disk (see the interpretation of the system below), and therefore we did not implement this method. As an additional measure, we have repeated the analysis of lines by Gaussian fitting and no substantial differences in our results were notable.  

Further discussion on spectral behavior of the object is presented below in Section\,\ref{sec:conclud}.  

\subsection{X-ray observation}
\label{sec:xray}

The {\sl Swift} observations of \sdss\ in November 2012  were obtained  as a target of opportunity (ObsID 00045632006),  prompted by the suspicion that the object is a magnetic CV \citep{2011AJ....142..181S}.  The XRT data were processed with  standard pipeline procedures. A 10\,ks long exposure showed no detectable signal from the object ($\le 0.005$ cnts/s).

\section{The orbital period}
\label{sec:porb}

The orbital period of \sdss\ was determined on the basis of  all acquired light curves ($V$  band),  and H$_\beta$ RV  variations by means of Discrete  Fourier Transform (DFT), as implemented in  {\sl {Period04}}\footnote{http://www.univie.ac.at/tops/Period04/}. The resulting power spectra  are presented in Figure\,\ref{fig:power}.  There appears to be one dominant  frequency in  power spectra of  photometric and radial velocity data sets. 
As usual, they are contaminated with  alias periods caused by observational cycles and frequencies of data measurement.  Although the peak frequency was obvious, we analyzed  the RV data  for  periods using 
{\it CLEAN} \citep{1978A&A....65..345S}, a DFT method, which convolves the power spectrum with the spectral window to eliminate alias 
frequencies originating from uneven data distribution.   The peaks in the power spectra of photometric and radial velocity data  closely coincide and correspond to the orbital period of the system.  

\begin{figure*}
\setlength{\unitlength}{1mm}
\resizebox{11cm}{!}{
\begin{picture}(100,70)(0,0)
\put (0,0)  {
 \includegraphics[width=15cm, bb=20 145 580 520, clip]{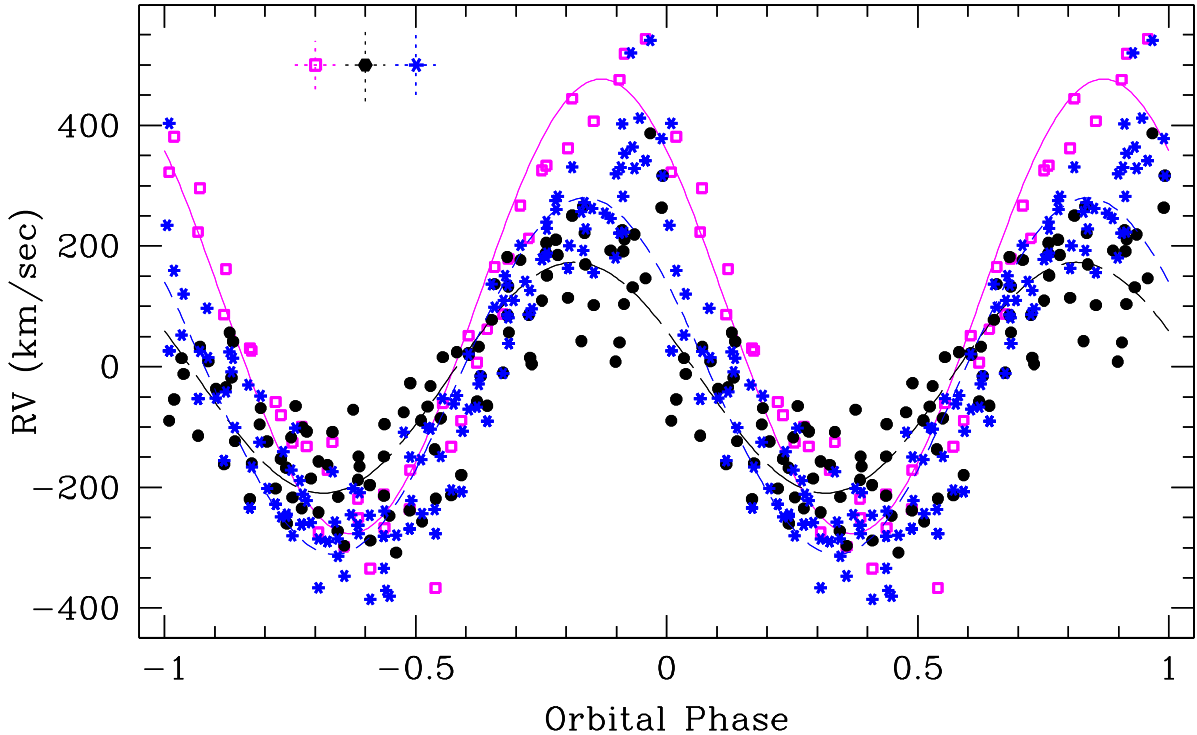}
   }
\end{picture}}
  \caption{The radial velocity curves  of \sdss.  The solid black circles are measurements of  the He\,{\sc ii} line and the black long-dashed line is a sinusoidal fit to the data. The blue stars and blue dashed line correspond to  H$_\beta$ and the magenta open squares and solid line are measurements and fit for the H$_\alpha$ line. All three lines show different semi-amplitude of radial velocities and phase shifts relative to each other as can be seen also from Table\,\ref{tab:rvs}. Phase zero corresponds to the eclipse. Two orbital periods are repeated for better presentation.}
  \label{fig:rv}
\end{figure*}

\begin{figure*}[t]

\setlength{\unitlength}{1mm}
\resizebox{11.5cm}{!}{
\begin{picture}(100,60)(0,0)
\put (-10,0)  {    \includegraphics[width=7.0cm,  clip]{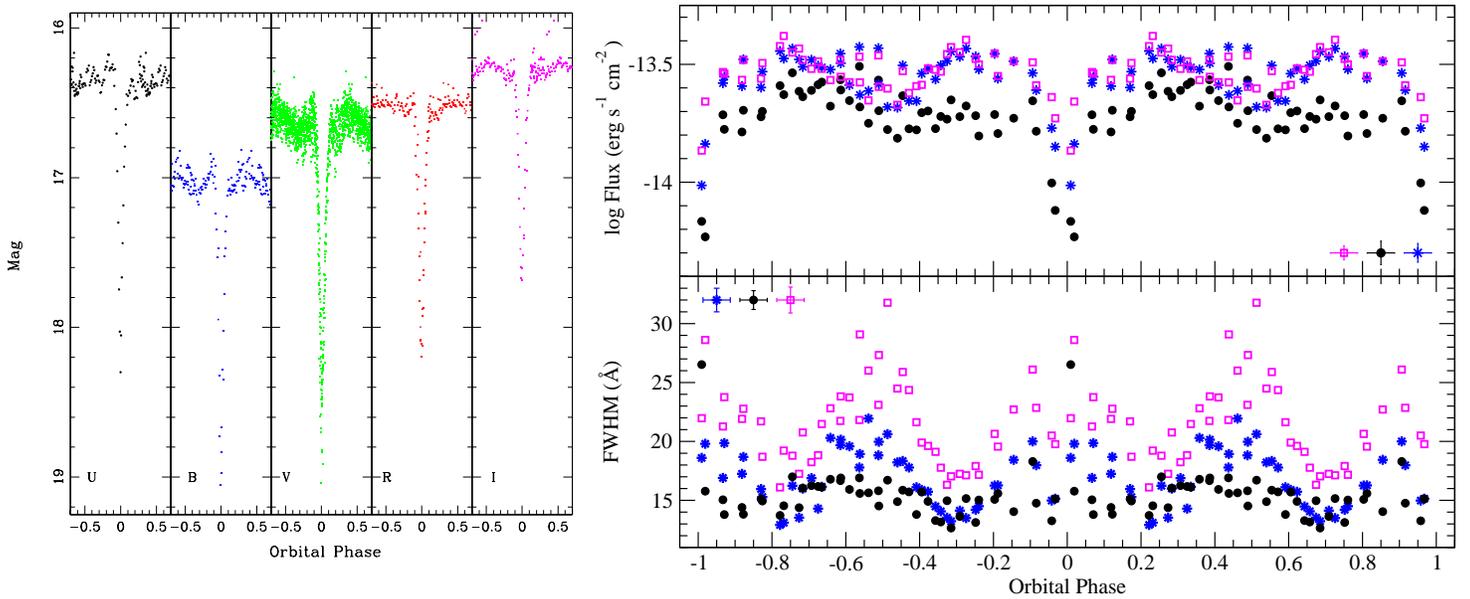}}
\put (60,-2)  {\includegraphics[width=10.1cm,  clip]{tovmassian_fig05b}}
\end{picture}}
\caption{UBV$RI$ light curves of  \sdss \,(left). Line parameters of  \sdss \,(right).  In the upper panel,  the fluxes of He\,{\sc ii} (black), H$_\beta$ (blue) and H$_\alpha$ (magenta) emission lines are plotted versus orbital phase.  In the lower panel, full width at half maximum measurements of  same lines are presented. The Balmer lines show a much larger broadening around phase 0.5 than  He\,{\sc ii}. At the same time, He\,{\sc ii}  shows deeper eclipses than the lower excitation lines.  For the sake of clarity, in the right panel the data are repeated to cover two orbital periods. }
  \label{fig:ubvri}
\end{figure*}

\begin{table}[t]
 \centering
    \caption{Radial velocity fit parameters.} 
\begin{tabular}{l|ccc} \hline
Line ID  &   $\gamma$  & Velocity &   Phase shift         \\ 
           &    km sec$^{-1}$         &     km sec$^{-1}$    &    $\phi$$^\dagger$  \\    \hline
H$_\alpha$   & $100.2 $    & $376.8$ &  0.88       \\
H$_\beta$  &   $-15.7$    & 295.4  &  0.91   \\
He\,{\sc ii}$^{*}$     &   $-18.3$    &   $191.3$  &   0.93  \\
\hline
\end{tabular}
\begin{tabular}{l}
$^{*}$  $\lambda 4686$\,\AA \\
$^\dagger$ P$_{orb}$ and HJD$_0$ determined from the eclipse   \\
ephemeris.
\end{tabular}
\label{tab:rvs}
\end{table}

The ephemeris deduced from the period analysis  is
$$ {\rm HJD}= 2\,455\,958.59184 + 0\fd1369745(4) \times {\rm E},  $$
where $\mathrm{HJD}_0$  refers to the eclipse center\footnote{  (see the reference to the program for uncertainties estimate)}. 
We folded all of the $V$-band data with the orbital period and obtained results consistent with the curves presented in Figure\,\ref{fig:lc} (with only slight differences outside eclipse.
The RV curves  of H$_\alpha$, H$_\beta$, and He\,{\sc ii},  folded with the orbital period, are presented in Figure\,\ref{fig:rv}.  Two thirds of spectral observations were made with the higher spectral resolution mode, but there is no notable difference in radial velocity measurements between the two sets. Least-square sinusoidal fits to the data for each emission line were calculated in the form of 
 $$RV(t) = \gamma+K_{\mathrm em}\times \sin(2\pi~ t/P + \phi),$$ where $\gamma$\  is the systematic velocity, $K_{\mathrm em}$ is the semi-amplitude 
of radial velocity and $\phi$ is the zero-phase.  The resulting values of $\gamma$, $K_{\mathrm em}$  and $\phi$ are presented in Table\,\ref{tab:rvs}.

\section{Phase dependent parameters}
\label{sec:param}

The eclipse profiles of SW\,Sex stars are usually non-symmetric and vary in shape with time \citep{1991AJ....102..272T}. 
In the case of \sdss\ the profile has stayed fairly stable over the course of our two year monitoring.
The light curve, outside the eclipse, is relatively flat. There is a small hump that appears immediately before the eclipse  and a depression appears after it , as the system emerges from it (see  Figure\,\ref{fig:lc}). The brightness then increases, peaking around $\phi=0.4$, and experiences another depression around 
$\phi=0.6-0.7$. This behavior   is seen, generally, in all other eclipsing SW\,Sex stars \citep{1991AJ....102..272T,1998MNRAS.294..689H}. In addition to the long series of $V$ photometry, we obtained multi-colour UBV$RI$ light curves of subsequent orbital cycles (see  Figure\,\ref{fig:ubvri},~left).  
Apparently, the pre-eclipse hump is more prominent in the blue  end of the optical spectrum (B \& V bands), while it is almost nonexistent in the $I$-band.  At the same time, the eclipse depth decreases to the red.  We can assume that, in eclipse, a hot source of continuum light is getting blocked by the secondary star, while the disk is not eclipsed entirely, and some cooler parts contributing in $R$ and $I$ bands are always visible. 
More telling are the variations of the emission lines parameters presented on the right side of Figure\,\ref{fig:ubvri}. In the upper panel, the intensity variation of H$_\alpha$, H$_\beta$ and He\,{\sc ii} lines are presented. The depth of eclipse increases with the excitation level of the line. H$_\alpha$ is obscured a little; its flux drops almost as much as in the counter-eclipse phase. H$_\beta$ is probably produced in a more compact area  and its eclipses are deeper. The high excitation line He\,{\sc ii} declines the most.  The semi-amplitude of the He\,{\sc ii} RV  is also  significantly smaller than of the Balmer lines (Figure\,\ref{fig:rv}). The orbital variability of the emission line full widths (FWHM) is shown in the right, lower panel in Figure\,\ref{fig:ubvri}. Here, He\,{\sc ii}  is almost flat for most orbital phases, while the Balmer lines varies greatly; the lower excitation line becoming much broader both around the eclipse and half the orbital phase.  

\begin{figure*}[t]
 \setlength{\unitlength}{1mm}
\resizebox{12cm}{!}{
\begin{picture}(100,100)(0,0)
\put (10,60){\includegraphics[width=70mm, bb=20 30 500 420, clip=]{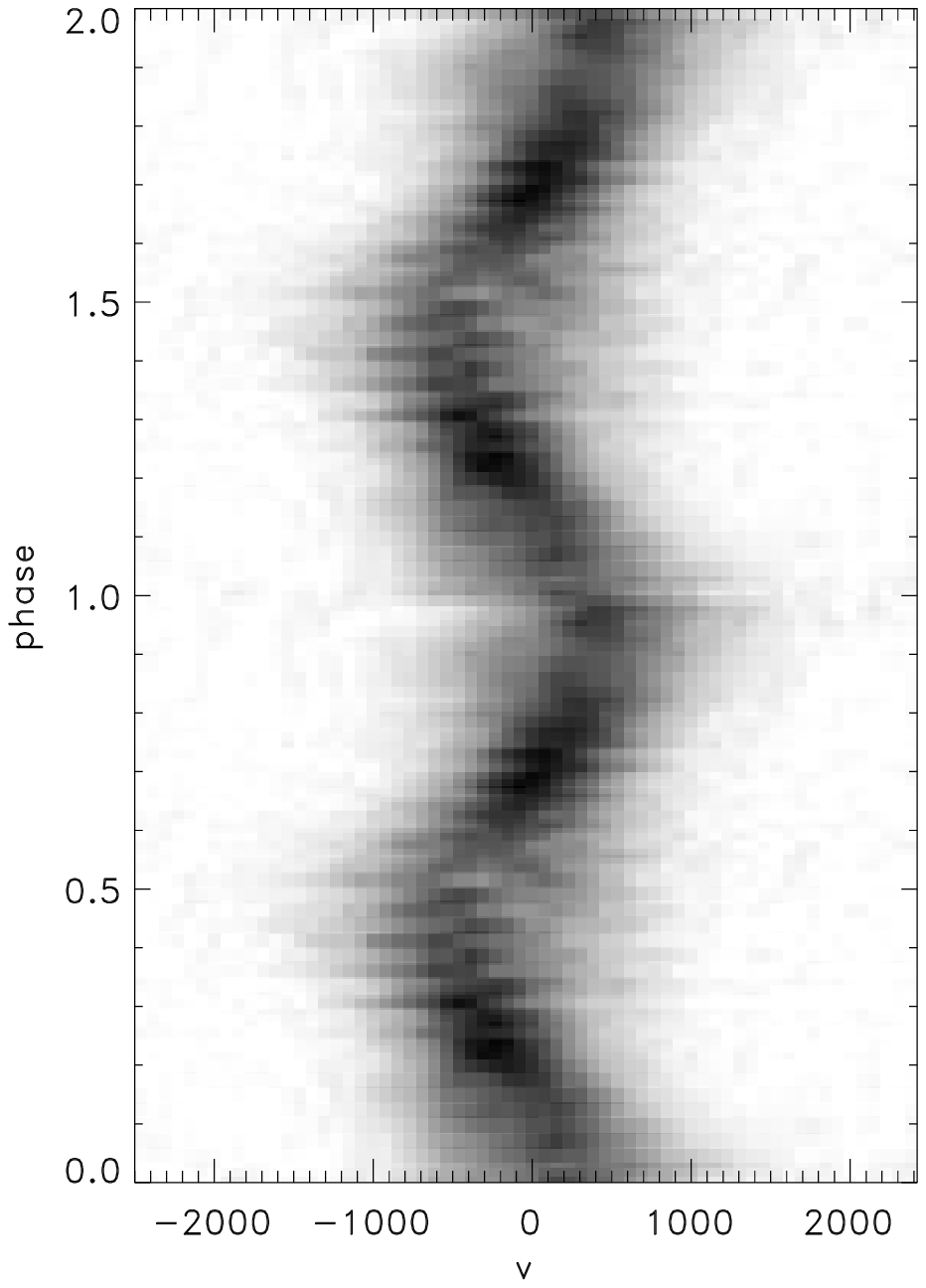}}
\put (50,60){\includegraphics[width=70mm, bb=20 30 500 420, clip=]{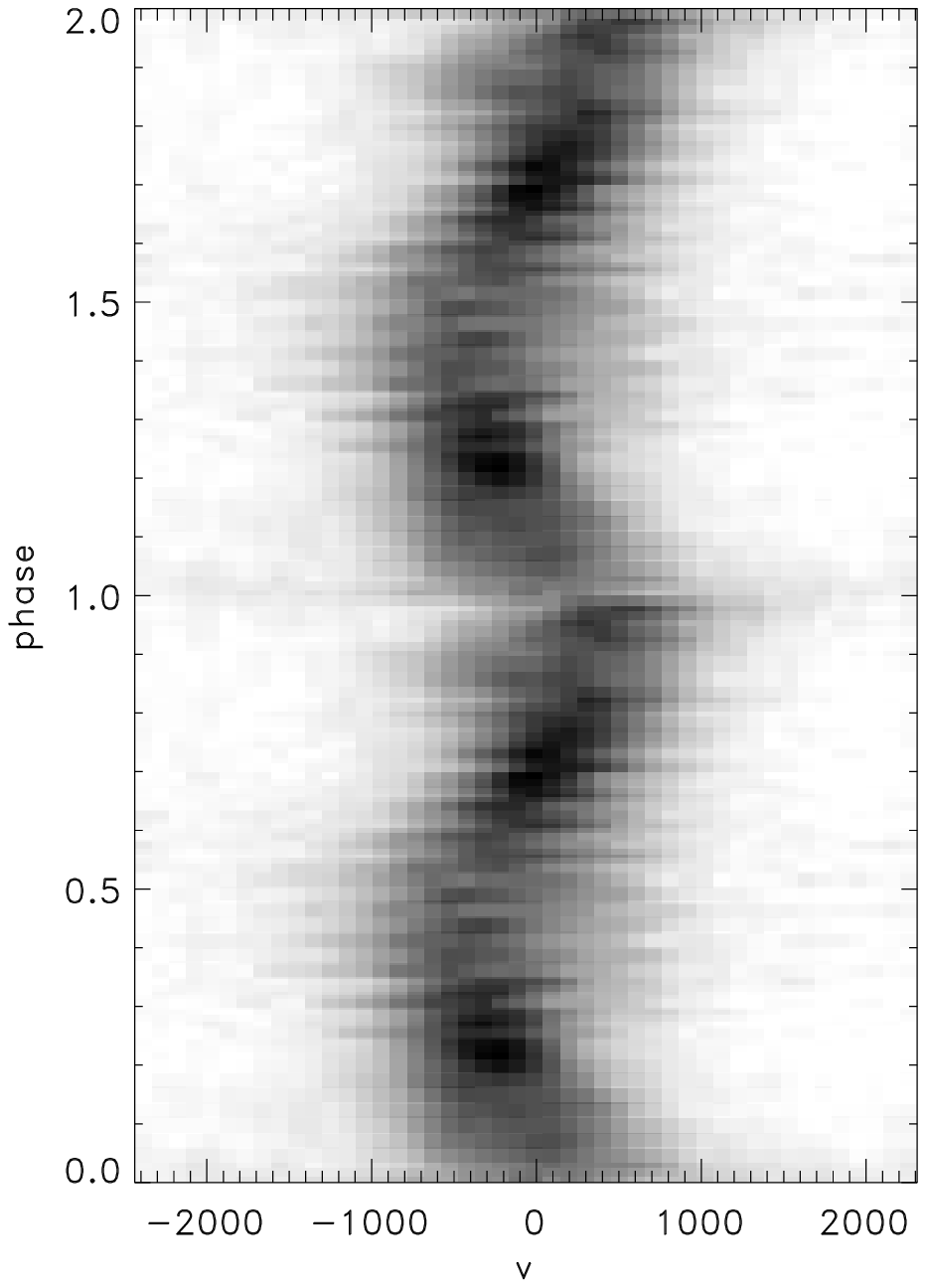}}
\put (90,60){\includegraphics[width=70mm, bb=20 30 500 420, clip=]{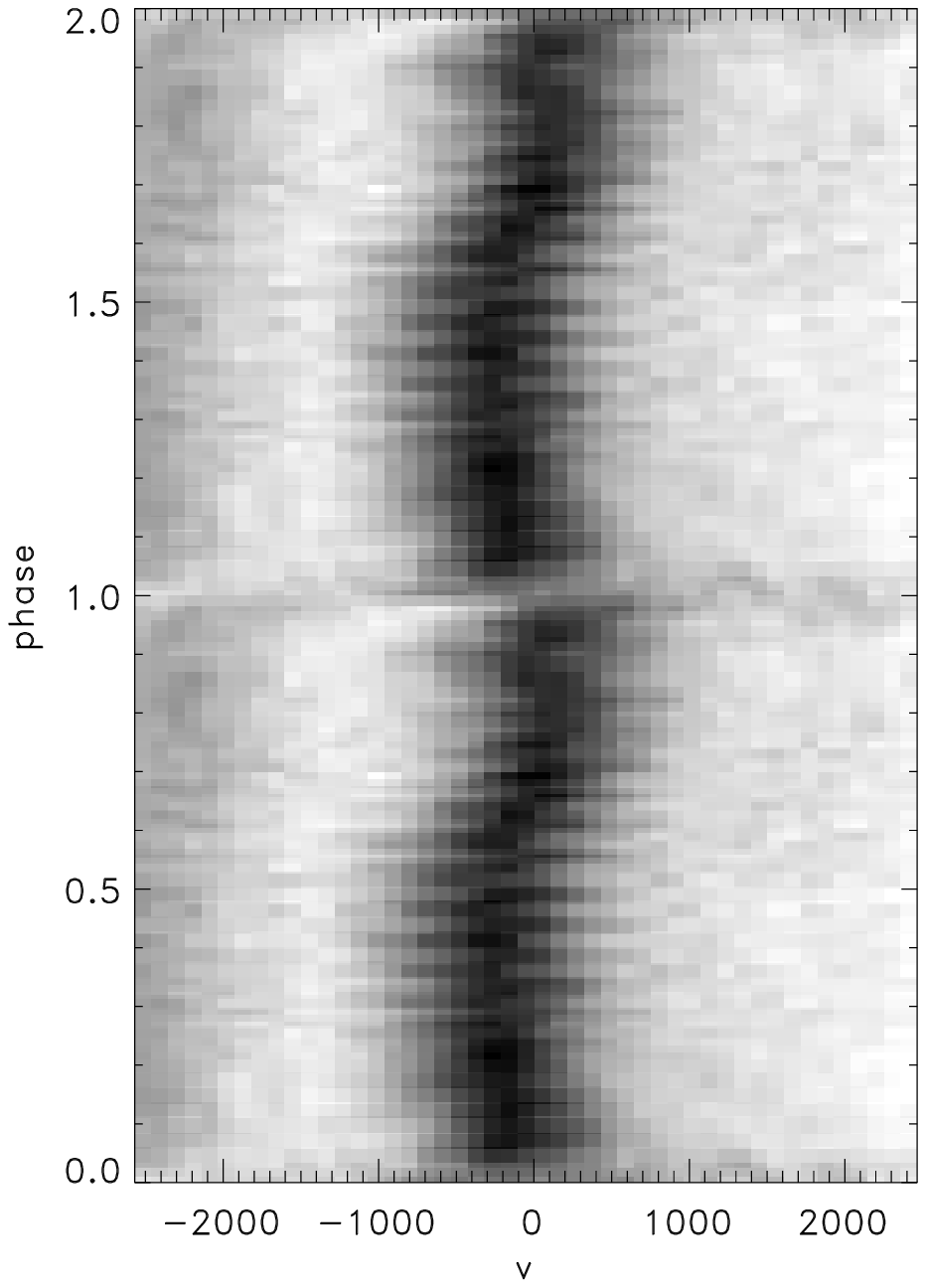}}
\put (10,0){\includegraphics[width=70mm, bb=20 30 500 420, clip=]{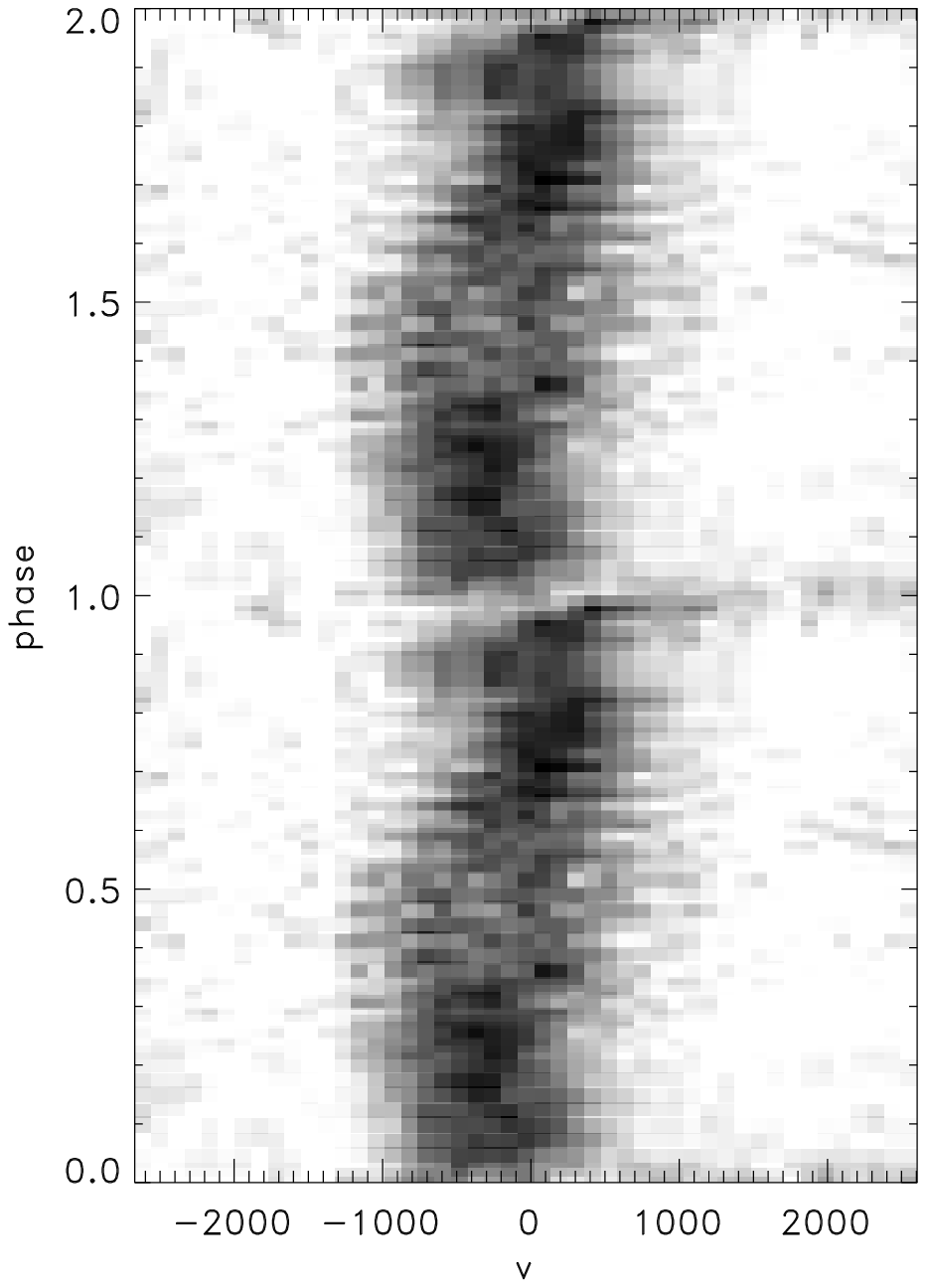}}
\put (50,0){\includegraphics[width=70mm, bb=20 30 500 420, clip=]{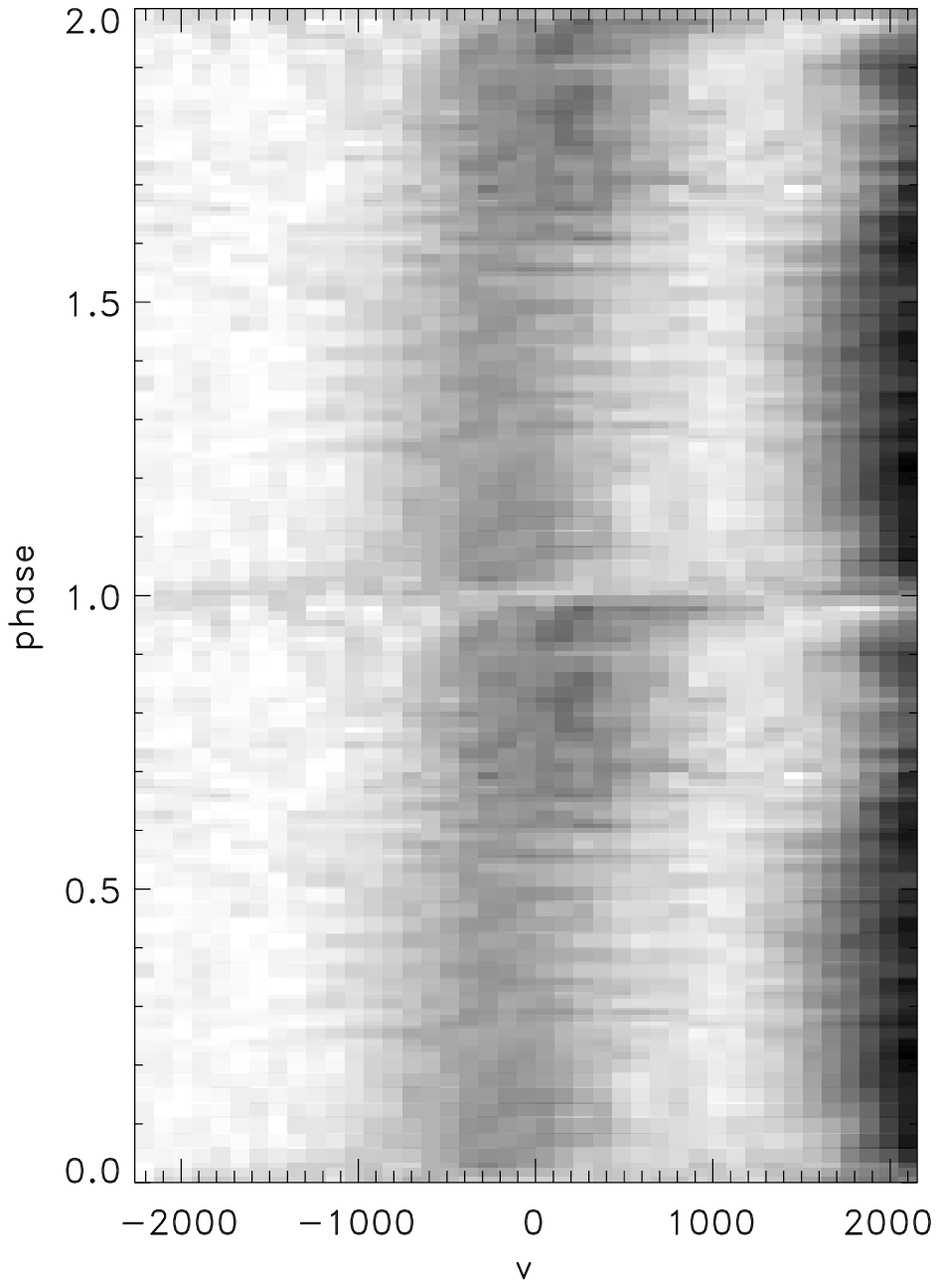}}
\put (90,0){\includegraphics[width=70mm, bb=20 30 500 420, clip=]{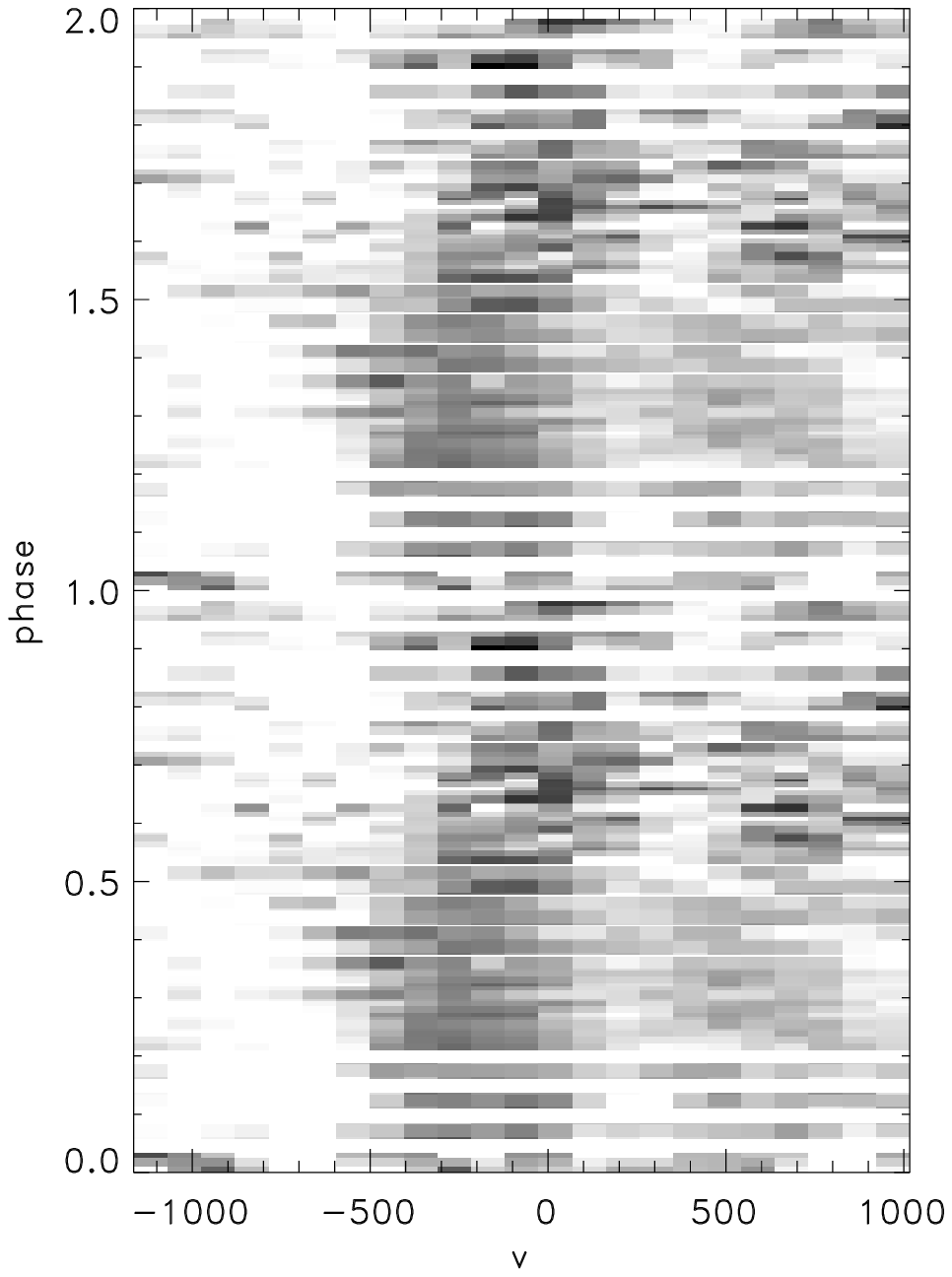}}
\end{picture}}
\caption{Trailed spectra  of selected lines of \sdss. 47 spectra  obtained on 6 and 7 February 2013 and covering slightly more than two orbital periods were used to construct the images. From the top left to the right in separate panels are presented  H$_\alpha$, H$_\beta$ and He\,{\sc ii}, and at the bottom also from the left to right:
He\,{\sc i}\,$\lambda 4471$\,\AA, complex of fluorescent lines  C\,{\sc iii}/N\,{\sc iii} and  Mg\,{\sc i} ( inverse gray scale, since the line is in absorption). }
\label{fig:trsp}
\end{figure*}

The behavior of H$_\alpha$, H$_\beta$, He\,{\sc ii},  He\,{\sc i}, fluorescent emission lines  C\,{\sc iii}/N\,{\sc iii} and  the Mg\,{\sc i}  absorption,  is depicted in a different manner by the trailed spectra 
in Figure\,\ref{fig:trsp}.  The trailed spectra were formed using relatively high S/N spectra obtained on two consecutive nights in February 2013. Among  the features that are  notable in the trailed spectra are:

\begin{itemize}
\item {right before the eclipse (starting from $\phi=0.8)$ there is a sharp zigzag in the  emission lines. }
\item {at phase $\phi=0$ (eclipse) there is a "tail" in emission extending to velocities exceeding 2000\,km/sec.}
\item {Balmer lines are narrower and appear more intense at phases $\phi=0.25$ and $0.75$ as demonstrated in the lower right panel of  Figure\,\ref{fig:ubvri}. }
\item {around phases $\phi=0.4 - 0.6$ emission lines become broader,  some He\,{\sc i} lines become double-peaked. There is flaring of lines at these phases, a feature common for SW\,Sex objects.}
\end{itemize}

Another common feature of SW\,Sex stars is the appearance in their spectra of the absorption lines not associated with the secondary. We can see that the    Mg\,{\sc i}  "b" line appears in almost  all phases, but becomes  more visible around phase $\phi=0.45\pm0.2$. 
It is also the phase when the emission lines become broader, probably due to a high velocity component and flaring. He\,{\sc i} line practically splits at that phase. 

\section{Discussion}

\label{sec:conclud}

This new CV, discovered by the SDSS, exhibits  all of the attributes necessary for it to be classified as a SW\,Sex star. It has a 3.29 hr orbital period placing it in the middle of the SW\,Sex stars period distribution, next to DW\,UMa, a classical representative of the class with which it shares many common features, as
outlined by \cite{2003ApJ...583..437A,2013MNRAS.428.3559D}.  Two other similar systems, 2MASS J01074282+4845188 and  HBHA 4705-03 have been discovered recently  \citep[][although the latter is not recognized as a SW\,Sex star]{2013A&A...551A.125K,2013AstL...39...38Y}, and are not yet included in the {\it big-list} of SW\,Sex objects compiled by Hoard\footnote{http://www.dwhoard.com/biglist}. But the number of CVs in the $\sim3-4$\,hours orbital range bearing characteristics of SW\,Sex stars grows,  and curiously, the fraction of  eclipsing systems among SW\,Sex type grows too, although we are far from understanding the phenomenon itself. 
Besides  the $V$-shaped  and non-symmetric eclipses,  \sdss\ shows high velocity  and single peaked emission lines forming a characteristic  "S-wave"  with phase shifts between different lines (Figure~\ref{fig:rv}).  
Even the lowest measured radial velocity amplitude encountered in He\,{\sc ii} ($\sim 200$km/s) is too high to be ascribed to the orbital velocity of a white dwarf in a CV with a three-and-half-hour orbital period. Such  systems would normally  consist of an 0.6--0.7\msun\ white dwarf and a $0.25\pm0.05$\,\msun\ secondary, with a maximum RV semi-amplitudes for the white dwarf and the secondary star  reaching  $\sim100$ and $\sim300$ km/sec, respectively.
Otherwise, the system should contain either an abnormally light-weighted white dwarf, or an unrealistically massive secondary star.

\begin{figure*}
\setlength{\unitlength}{1mm}
\resizebox{11cm}{!}{
\begin{picture}(100,110)(0,0)
\put (20,0)  {  \includegraphics[width=14cm,  bb = 25 180 580 810,clip]{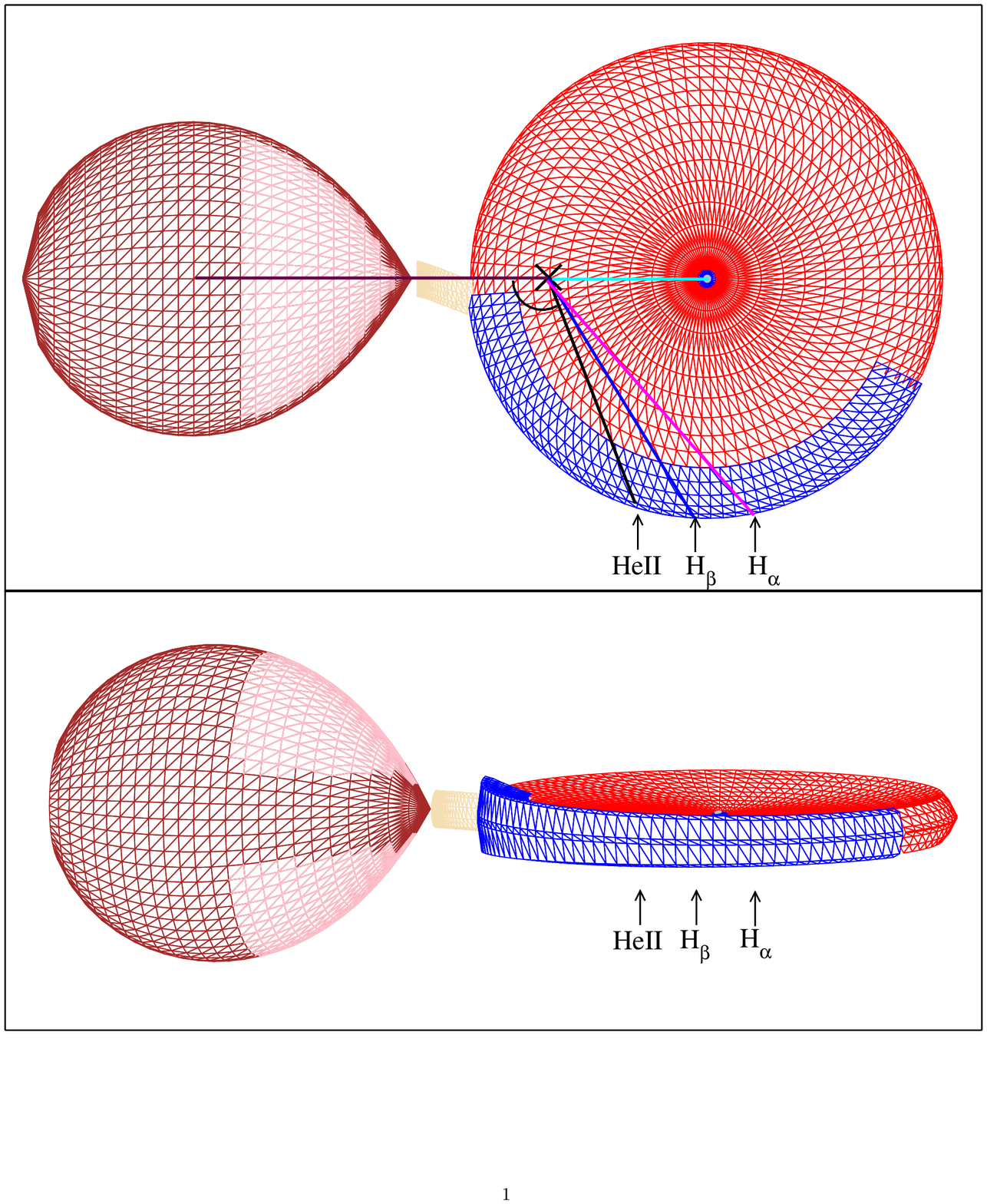}
   }
\end{picture}}
  \caption{SW Sex  model.  The parameters for this one are selected according to Table\,\ref{t:mod} to reproduce the light  and radial velocity curves closely resembling  the observations. The calculated curves are  presented in the following figures. The extended hot spot, or hot line is  drawn by blue. Marked are the weighted centers of  areas  where the corresponding emission lines are formed. Because of the temperature gradient in the arc and stratification of lines each of them revolves around the center of mass denoted by "x"  with different velocity and phase.  }
  \label{fig:mod}
\end{figure*}

The single-peaked emission lines remain most baffling features of SW\,Sex stars. Most authors  employ models that help to elevate the ionized matter above the orbital plane to explain the formation of such line profiles,  i.e bright-spot overflow plus accretion disc wind \citep{1986ApJ...302..388H,1994MNRAS.267..153H,1995MNRAS.277..777D}, magnetic accretion column \citep{1989AJ.....97.1752W}, or
magnetically driven outflow \citep{1993MNRAS.265L...5T, 1995MNRAS.275....9W}.  These maybe not be necessary,  if we take a clue from  the study of BT\,Mon, an old nova with a full range of SW\,Sex observational characteristics, by \citet{1996ApJ...456..777W}. These authors show that the emission lines originate where the bright spot should be (or slightly further), yet they claim they do not see the presence of an accretion disk.  Let us keep in mind that SW\,Sex stars are high mass transfer rate systems and they have steady hot disks similar to those of dwarf novae in outburst \citep{1990ApJS...73..441S}. Such disks are optically thick and maintain a delicate balance of absorption and emission lines, often appearing almost featureless. We may assume, that in \sdss\ and other SW\,Sex objects, the accretion disk has a large contribution in the continuum but little in the lines, and that the bright spot is either the only or the  dominant source of the emission lines.   Thus, the observer of spectral lines "does not see" the presence of the accretion disk and detects only single peaked emission lines from an  area known as the bright spot. This can be exacerbated by the elongated down the stream bright spot and the geometrical thickness of disk edges  shadowing the emission from its surface.

In fact \citet{1997MNRAS.291..694D}  proposed the exact same idea; they even  considered  an extended bright spot. But they stopped short affirming presence of such structure in the disk citing an  absence of detailed hydrodynamical calculations describing it.    
Recently, \citet{2008ARep...52..318B,2012A&A...538A..94K} performed such three-dimensional gas-dynamical simulations for accretion disks.  Their calculations   show   the flow of matter  in SS\,Cyg, a dwarf nova with a  \mdot $\sim8\times10^{-9}$\,\msun/year mass transfer rate, a  typical value for dwarf nova.  One would expect that the effects obtained in that simulation would get more important with an increase of mass transfer rate.  According to their calculations, apart form the stream of matter  feeding the  accretion disk  from the secondary star, there is an outflow of matter leaving the accretion disk and forming a circum-disk halo. The former apparently plays an important role in SW\,Sex stars as in all interacting binaries. In this case, its role is enhanced by the extension of the spot into what  \citep{2008ARep...52..318B}  call a "hot line" or arc. Consideration of the latter, i.e. the outflow of matter form the disk,  can help to explain the additional features.  We can see, in the right panel of Figure\,\ref{fig:ubvri}, that in phase $\phi=0.5$ the Balmer lines are twice broader than at other phases, but their intensity decreases.  They also show irregular variation of width, also called line flaring.  This can be explained by the presence of the circum-disk halo which produces lines  with broader  velocities. It  can also be responsible for the appearance of the absorption features. At $\phi=0.25$ and $\phi=0.75$ both streams are viewed perpendicular, thus the dispersion of the velocities is smaller and the lines narrower.  During the eclipse, the halo remains mostly visible,  preventing a total  eclipse of the low-excitation lines.
The system brightens at  $\phi=0.3 -0.4$ and exhibits irregular variability,   which \citet{2013MNRAS.428.3559D} explain as  a non-uniform disc edge. Their model does not explain why it is non-uniform, or why the rim on that side of the disk is not as wide.  \citet{2004ApJ...615L.129K} came to a similar conclusion using UV spectra, noting a presence of absorbing material at the phase opposing the eclipse and its nonuniform physical conditions. However if there is outflow, then it will probably introduce a necessary  non-uniformity to the disk edge.
The flaring of the emission lines has been interpreted as an evidence of magnetic accretion \citep{2001ApJ...548L..49R,2002MNRAS.337..209R,2009MNRAS.395..973R}. The claim is supported by evidence of periodic nature of flaring, and in some cases, a more substantial indication is the detection of circular polarization. We do not question presence of magnetic white dwarfs in SW\,Sex stars and truncation of the accretion disk, but we do not have any manifestation of magnetic field in case of \sdss. We analyzed the equivalent width variability of H$_\beta$ and He\,{\sc {ii}} lines and found no periods except the orbital and an alias related to the  exposure length of individual spectra. We also believe that a magnetic accretion modulating the emission lines width would produce substantial X-ray emission, but we detected none.  

\begin{figure*}[t]
\setlength{\unitlength}{1mm}
\resizebox{12cm}{!}{
\begin{picture}(100,80)(0,0)
\put (-10,-5)  {
   \includegraphics[width=14.5cm,  bb =10 180 600 600,clip]{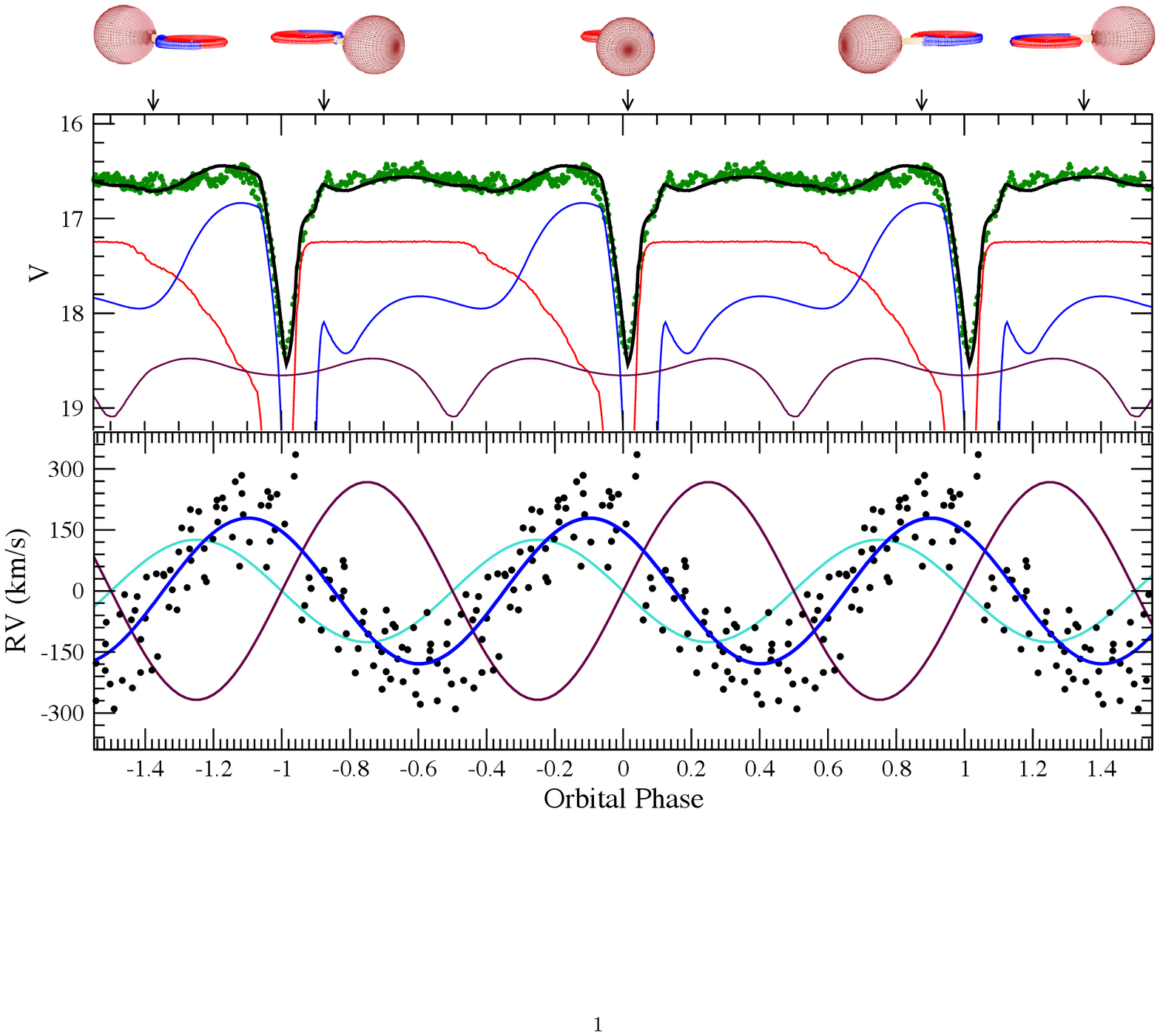}
   }
\end{picture}}
  \caption{Light and radial velocity curves calculated from the model. At the top of the figure five different perspectives of the model are presented. Their corresponding phases are marked by ticks on the outer side of the upper panel. Phase 0.0 in this figure corresponds to the inferior conjunction, unlike previous figures where it corresponded to the eclipse center ($\phi=0.025$ difference). In the upper panel the computed  V-band light curves of the disk (red), the hot spot (blue) and the secondary star (brown) are plotted along the sum of all components,  presented as a black line. The green points  are observations in  V-band. The contribution of white dwarf is negligible and lays out of the figure. In the bottom panel the computed radial velocity of stellar components are presented. The higher velocity brown curve corresponds to the secondary star and the cyan to the white dwarf. They are not observed. The points correspond to the observed radial velocity of   He\,{\sc ii} line and the blue curve is the computed velocity from the model. Two Balmer lines marked in  Figure\,\ref{fig:mod}  are not presented, not to complicate the plot, but they  describe the observed points as good as the  He\,{\sc ii}.   }
  \label{fig:modPhot}
\end{figure*}

\section{The model}
\label{sec:model}

We developed a  geometric  model  of a SW\,Sex system, presented in Figure\,\ref{fig:mod}, to reproduce the observed photometric and spectral features  of \sdss. The system is comprised of a primary white dwarf, a secondary red dwarf star; a stream of accretion matter, a thick  $z^d(r) = z^d(r_{out}) (r/r_{out})^\nu$ accretion disk, and  an elongated and extended hot spot/line.   The white dwarf is a sphere, defined by the mass-radii relation in Warner \citep[2.83b,][]{1995CAS....28.....W}. The secondary is assumed to fill its Roche lobe, and the Roche lobe shape is directly calculated using  equation 2.2 \citep[][]{1995CAS....28.....W}  for equipotential $\Phi$(L$_1)$.
The mass and temperature of the primary  are defined as initial parameters. The mass  of the secondary is obtained from the 
mass-period relation 2.100 \citep[][]{1995CAS....28.....W}, which also defines the effective temperature of the secondary, using relations $\log T^{\mathrm {eff}}_2 \propto \log M_2$ from \citet{2007MNRAS.382.1073M}.
 The  illumination of the secondary by the primary is also included. A standard  temperature gradient $$T^d(r) \sim T_* \times r^{-3/4},$$ where $$T_* = \left(\frac{3GM_{wd}\dot{M}}{8\pi \sigma R_{wd}^3}\right)^{1/4}$$  \citep[see 2.35, 2.36,2.37][]{1995CAS....28.....W} between the inner and the outer edges ($r_{out} \approx 0.60 a /(1+q)$) of the disk  was assumed.
 The impact of the stream with the outer rim of the disk   forms a shock-heated trailing arc   along  the rim  as was described by  the schematic model by \citet[][Fig.12 therein]{2003AJ....126.2473H} and the hotline defined in \citet{2008ARep...52..318B}.  The vertical size $z^{\mathrm s}$ and the temperature $T^{\mathrm s}$ of  the arc-like hot spot are  not uniform; it is wider and  hotter at the impact point ($\varphi_{min}$) and  declines both in temperature and vertical extension, according to an arbitrarily selected manner described by the following equations:
 \[ T^{s}(\varphi)=T^{d}(1.0+\gamma^{T}f(\varphi)^\frac{3}{2}) \]
  \[ z^{s}(\varphi)=z^{d}(1.0+\gamma^{z}f(\varphi)^\frac{5}{2}), \]
 where $\varphi$ is  the angle between the line connecting the stars in conjunction and the direction of the spot viewed from the center of mass; $f(\varphi) \equiv a\times\varphi+b$ is a linear function with constants $a$ and $b$ defined by boundary conditions of $f(\varphi_{min})  = 1$; and  $f(\varphi_{max})  = 0$, and  $\gamma^T,\gamma^z$ are  free parameters. 
 
The surface of each component of the system is divided in a series of triangles  as shown in Figure\,\ref{fig:mod}. We assume that  each triangle emits as a blackbody with corresponding  temperature. Figure\,\ref{fig:modPhot} presents  geometry snapshots (top) and the model calculations of light and radial velocity curves (bottom two). 
\begin{table}[t]
 \caption{Model parameters used to compute the photometry and radial velocity curves.} 
\begin{tabular}{ll}                                  \hline
{\it System input:}                           &				\\
Period                                &   11834.6 sec          \\
Mass of Primary	         &    0.6$M_\odot$       \\ 
mass ratio          	 	 &    0.47                      \\    
system inclination               &    84$^o$                 \\   
radial disk size $r_{out}$                  &	  $0.40 R_\odot$  \\
vertical disk size $z^d(r_{out})$  &          0.10 $R_\odot$ \\
$\nu$    radial slope                                     &  3 \\
$\dot{M}$			&  $1.2\times10^{-10}M_\odot/yr$ \\ 
Distance &  290 pc \\ \hline
{\it Hot spot/line:}  &     \\
width    &     0.2$\times r_{out}$ \\
length   & 140$^o$ \\   
$\gamma^T$ & 0.75  \\
$\gamma^z$ & 0.75 \\ \hline
{\it System output:} & \\
Mass of Secondary & 0.28$M_\odot$ \\
$T_2$ &    3250 $K$ \\
 $a$ separation & 1.07 $R_\odot$ \\

\end{tabular}

\label{t:mod}
\end{table}
The light curves of  individual components of the binary system were  obtained by integrating the emission from all the elements lying  in a sight of view and are presented in the middle panel of  Figure\,\ref{fig:modPhot}.  The flux from the white dwarf is almost totally blocked by the thick disk. Two equally dominant sources of emission from the model light curves are the accretion disk and the bright spot/arc (denoted by the red and blue lines respectively). The disk's contribution is more significant after the eclipse, when the bright spot is largely hidden behind it.  In contrast,  the form of the light curve before the eclipses, is mostly determined by the bright spot. The eclipse corresponds to the blockage of the bright spot by the secondary.   Thus, the observational phase $\phi=0$, which is assigned to the conjunction of the stellar components, does not coincide with the zero phase in Figure\,\ref{fig:modPhot}.  The fit is not optimized mathematically, but it is sufficient to explain all the observed features: the eclipse, the pre-eclipse-hump, the depression around  phase 0.6-0.7, and even the little wiggle at  $\phi=0.18$  reasonably well. The main parameters of the model are given in Table\,\ref{t:mod}. We can achieve satisfactory qualitative resemblance between the modeled light curves and the observations by simply varying the system inclination and the hot spot parameters.  All  other parameters are fixed.

The model describes also quite well the RV amplitude and phase of the emission lines as basically emanating  from the hot spot. Since the hot spot is extended, and its temperature is the highest at  the impact point, there is a stratification of temperatures and hence the excitation levels along the arc. The lines of different excitation energies are predominantly formed  at a different angles from the mass center, as marked in Figure\,\ref{fig:mod}. He\,{\sc ii}  arises from an area located the closest to the the center of mass of the binary and thus, has the lowest amplitude of the radial velocity.  The lower the excitation level, the further  is the weighted center of the line  forming region along the arc, and, therefore, the larger is the radial velocity amplitude. The phasing of the lines relative to the orbital phase changes accordingly. In the bottom panel of Figure\,\ref{fig:modPhot} the calculated radial velocity curve for He\,{\sc ii} line is presented and it excellently  describes points obtained from the observations. Also shown are calculated velocities of stellar components, which are not observed. Calculated velocities of Balmer lines are not presented, so as not to crowd the figure.  
 
 The model also is  able to describe fine details, in particular, what happens at  the phases $\phi=0.9$ to $1.0$. The white dwarf (and the accretion disk in general) has already reached the maximum positive velocity and it is moving toward stellar conjunction with decreasing radial velocity (light blue  curve). But emission lines, formed predominantly in the bright arc still have an increasing radial velocity peaking around phase  $\phi=0.9$ (blue curve). Around that phase, however, the secondary starts eclipsing part of the bright arc where the matter with the growing velocities is concentrated, and the observer sees only  the far extreme of the arc, which moves in the opposite direction. Thus, for a brief moment before the total eclipse of the bright arc, we detect a less intense emission line with decreasing RV. This creates a zig-zag in the trailed spectra (see Figure\,\ref{fig:trsp}). The zig-zag is best pronounced in the He\,{\sc ii} line, since it is formed closer to the stream/dick impact point, which is eclipsed first, leaving rest of the arc visible.  At the depth of the eclipse ($\phi=0.025$ in phases  determined by radial velocities), the very tip of the retreating half of the disk or the hot arc is still visible (see the cartoon at the top of the Figure\,\ref{fig:modPhot}). It contains matter moving away from the observer with the sum of Keplerian and orbital velocities, and appears in the RV curves and trailed spectra as a high velocity  tail.   This model does not take into account the matter outflowing from the opposite side of the hot arc side in the disk. We think that the inclusion of that element would allow  fine tuning of the model to the observations.

\section{Conclusions}
\label{sec:conclud2}

We have unveiled a new eclipsing SW\,Sex star. It has an orbital period $3.287$\,hr,  at the centre of a range of periods where the great majority  of SW\,Sex stars cluster. It has all observational characteristics of this type of stars and, as such, increases the number of eclipsing systems of this particular type at these periods.   
We believe that this clustering is something not well understood and should be addressed, but it is beyond the scope  of this study.  On the other hand we have been able to reproduce the photometric and radial velocity curves of \sdss\ by simply assuming that the emission lines are predominantly formed/observed from the hot, extended spot, without the additional speculative mechanisms proposed in other models. We think that the absence of lines from other parts of the disk is a consequence of a disk having a  high temperature and a surface density regime, and partially due to self obscuration by the thick rim at the same time.   We also, believe  that there is an outflow of matter from  the opposite to the arc side of the disk, where the absorption features observed in SW\,Sex objects are formed. That outflowing matter may also contribute to the high velocity components of the low-excitation lines.  

\acknowledgments  

MSH and DGB is grateful to CONACyT for grants allowing their post-graduate studies.  
GT and  SZ acknowledge PAPIIT grants IN-109209/IN-103912  and CONACyT grants 34521-E; 151858 for resources provided toward this research. We would like to thank the {\sl Swift} team for opportunity to observe the object and their prompt response to our application. Our thanks to Dr. J. Echevarria for reading the article and for his critical and helpful comments.

\end{document}